\documentclass[aps,pra,superscriptaddress,twocolumn,10pt,longbibliography]{revtex4-1}

\usepackage{amsmath}
\usepackage{graphicx}
\usepackage{bm}
\usepackage{color}

\date{\today}

\newcommand{\dd}{\mathrm{d}}
\newcommand{\rpsi}{|\psi \rangle}

\definecolor{darkblue}{rgb}{0.1,0.2,0.6}
\definecolor{darkred}{rgb}{0.8,0.1,0.2}
\usepackage[colorlinks,citecolor=darkblue,linkcolor=darkred,urlcolor=darkblue]
{hyperref}
\usepackage[all]{hypcap}

\usepackage{etoolbox}
\usepackage{mathtools}
\usepackage{amsmath}%
\usepackage{amsfonts}%
\usepackage{amssymb}

\begin{document}

\title{Many-body localization and delocalization in large quantum chains}
\author{Elmer~V.~H. Doggen}
\email[Corresponding author: ]{elmer.doggen@kit.edu}
\affiliation{Institut f\"ur Nanotechnologie, Karlsruhe Institute of Technology, 76021 Karlsruhe, Germany}
\author{Frank Schindler}
\affiliation{Department of Physics, University of Zurich, Winterthurerstrasse 190, 8057 Zurich, Switzerland}
\author{Konstantin~S. Tikhonov}
\affiliation{Institut f\"ur Nanotechnologie, Karlsruhe Institute of Technology, 76021 Karlsruhe, Germany}
\affiliation{L.~D. Landau Institute for Theoretical Physics RAS, 119334 Moscow, Russia}
\author{\mbox{Alexander~D. Mirlin}}
\affiliation{Institut f\"ur Nanotechnologie, Karlsruhe Institute of Technology, 76021 Karlsruhe, Germany}
\affiliation{L.~D. Landau Institute for Theoretical Physics RAS, 119334 Moscow, Russia}
\affiliation{\mbox{Institut f\"ur Theorie der Kondensierten Materie, Karlsruhe Institute of Technology, 76128 Karlsruhe, Germany}}
\affiliation{Petersburg Nuclear Physics Institute, 188300 St.~Petersburg, Russia}
\author{Titus Neupert}
\affiliation{Department of Physics, University of Zurich, Winterthurerstrasse 190, 8057 Zurich, Switzerland}
\author{Dmitry~G. Polyakov}
\affiliation{Institut f\"ur Nanotechnologie, Karlsruhe Institute of Technology, 76021 Karlsruhe, Germany}
\author{\mbox{Igor~V. Gornyi}}
\affiliation{Institut f\"ur Nanotechnologie, Karlsruhe Institute of Technology, 76021 Karlsruhe, Germany}
\affiliation{L.~D. Landau Institute for Theoretical Physics RAS, 119334 Moscow, Russia}
\affiliation{\mbox{Institut f\"ur Theorie der Kondensierten Materie, Karlsruhe Institute of Technology, 76128 Karlsruhe, Germany}}
\affiliation{A.~F. Ioffe Physico-Technical Institute, 194021 St.~Petersburg, Russia}

\begin{abstract}
 We theoretically study the quench dynamics for an isolated Heisenberg spin chain with a random on-site magnetic field, which is one of the paradigmatic models of a many-body localization transition. We use the time-dependent variational principle as applied to matrix product states, which allows us to controllably study chains of a length up to $L=100$ spins, i.e., much larger than $L \simeq 20$ that can be treated via exact diagonalization.  For the analysis of the data, three complementary approaches are used: (i) determination of the exponent $\beta$ which characterizes the power-law decay of the antiferromagnetic imbalance with time; (ii) similar determination of the exponent $\beta_\Lambda$ which characterizes the decay of a Schmidt gap in the entanglement spectrum, (iii) machine learning with the use, as an input, of the time dependence of the spin densities in the whole chain. We find that the consideration of the larger system sizes substantially increases the estimate for the critical disorder $W_c$ that separates the ergodic and many-body localized regimes, compared to the values of $W_c$ in the literature. On the ergodic side of the transition,  there is a broad interval of the strength of disorder with slow subdiffusive transport. In this regime, the exponents $\beta$ and $\beta_\Lambda$ increase, with increasing $L$, for relatively small $L$ but saturate for $L \simeq 50$, indicating that these slow power laws survive in the thermodynamic limit. From a technical perspective, we develop an adaptation of the ``learning by confusion'' machine learning approach that can determine $W_c$.
\end{abstract}

\maketitle

\section{Introduction}

Many-body localization (MBL) refers to the localization of particles by disorder in the presence of interactions at non-zero energy density (for recent reviews, see \cite{Nandkishore2015a, Altman2015a, Abanin2017a}), as opposed to the conventional Anderson localization \cite{Evers2008a} which describes non-interacting particles in the presence of disorder.
MBL is of fundamental interest for understanding metal-insulator transitions and disordered superconductors.
From a numerical perspective, MBL is a notoriously difficult problem to describe because of its many-body nature, sensitivity to finite-size effects and the requirement for ensemble averaging over many realizations of disorder.
On a conceptual level, the study of MBL ties to the thermalization of quantum systems and the bridge between microscopic dynamics and quantum statistics \cite{Rigol2008a}. 

The study of MBL did not begin in earnest until the development of landmark theories \cite{Gornyi2005a, Basko2006a} predicting a temperature-driven transition to a localized phase, now known as an MBL phase.
At the same time, advances in the increase of computational power and the development of new algorithms based on tensor networks like matrix product states (MPS) have dramatically accelerated the numerical study of disordered interacting systems, albeit thus far restricted to ground states, short times, boundary-driven systems or very strong disorder \cite{Znidaric2016a, Khemani2016a, Yu2017a, Wahl2017a, Doggen2017a}.
Experimentally, MBL has been reported in ultracold atoms \cite{Schreiber2015a, Choi2016a}, trapped ions \cite{Smith2016a}, and dipolar spins in diamond \cite{Kucsko2018a}.
Analytically, the existence of an MBL region in the phase diagram of a disordered quantum spin chain has 
been proven based on an assumption about energy level repulsion \cite{Imbrie2016a}.

In this work, we investigate a Heisenberg chain with isotropic interactions and a random magnetic field, which has become one of the paradigmatic models for the investigation of the MBL-related physics. 
In particular, Ref.~\cite{Pal2010a} considered system sizes $L$ from $L= 8$ to $L=16$ and found crossing points in level statistics and relaxation of spin modulations plotted as a function of disorder $W$ (in notations of the present work) for various system sizes. {By analogy with the scaling analysis \cite{Shklovskii1993} of the Anderson transition in non-interacting systems, these crossing points may serve as an indication of the MBL transition \cite{Oganesyan2007}.} However, a rather strong drift of the crossing points was found, from $W\simeq 2$ to $W\simeq 3$, as $L$ increased from 8 to 16, making it difficult to locate the transition. In a later work  \cite{Luitz2015a}, systems of larger sizes, up to $L=22$, were investigated via exact diagonalization (for the current state-of-the-art, see Ref.\ \cite{Pietracaprina2018a}), and the estimate for the transition point $W_c \simeq 3.7$ was obtained, consistent with the upper bound $W_c \lesssim 4$ found in Ref.~\cite{Berkelbach2010a}. On the other hand, Ref.~\cite{Devakul2015a} used a numerical linked-cluster expansion for the entanglement entropy in the thermodynamic limit and obtained a lower bound $W_c \gtrsim 4.5$ which appears to be in conflict with the exact-diagonalization results quoted above. 

All in all, the current status of the numerically obtained results is still controversial; simulations of small systems can easily miss physics relevant at longer length scales even if ameliorated by finite-size scaling approaches.
Most importantly, the strong dependence of the apparent transition point on the system size calls for a detailed numerical study of physical observables characterizing the MBL transition in systems of size $L$ that is well beyond the reach of exact diagonalization. To achieve this goal, we consider the MPS-based approach, which allows us to controllably study the long-time quench dynamics around the MBL transition in spin chains of length up to $L=100$. We also apply machine-learning techniques to further analyze the data obtained from MPS. 

Clearly, not only the position of the MBL transition is of interest but also a more detailed insight into the physics of phases around it.  Previous works on one-dimensional (1D) systems indicated that a part of the delocalized phase adjacent to the transition is characterized by a slow, subdiffusive dynamics \cite{Serbyn2013b, BarLev2015a, Agarwal2015a,  Agarwal2016a, Luitz2016a}. However, it was also found \cite{Bera2017a}, within exact-diagonalization studies of systems with size up to $L=24$, that an apparent exponent characterizing this subdiffusive phase changes substantially with the system size. This poses a question of whether the subdiffusive behavior is a genuine property of the system in the $L \to \infty$ limit, or merely a transient feature which characterizes the relatively small systems. {Previously, in larger systems a transition from diffusive to subdiffusive behavior was reported at relatively weak disorder $W \approx 0.55$ \cite{Znidaric2016a}, however to date numerical results for large systems at stronger disorder are still lacking.} {It is therefore important to numerically study substantially larger systems in the crossover from the ergodic to the localized regime, which elucidates the nature of the MBL transition}. This is another motivation for the present investigation of the MBL physics in a Heisenberg chain within the MPS-based approach.

\section{Model and method}

\subsection{Disordered Heisenberg XXZ chain}

We consider the Heisenberg XXZ chain with an on-site random field on a lattice of length $L$ with open boundary conditions, as described by the Hamiltonian
\begin{equation}
 \mathcal{H} = \sum_{i=1}^L\left[ \frac{J}{2}\Big(S_i^+S_{i+1}^- + S_i^- S_{i+1}^+ \Big) + \Delta S_i^z S_{i+1}^{z} + h_i S_i^z\right]. \label{eq:hamiltonian}
\end{equation}
Here $S_i^+$,  $S_i^-$, and $S_i^z$ are the standard spin-$1/2$ Pauli operators corresponding to the site $i$,
and the on-site field $h_i$ takes random values according to a uniform distribution $h_i \in [-W,W]$.
In the following, we set $J, \hbar \equiv 1$ as a choice of units, and put $\Delta = 1$, i.e., we consider the isotropic Heisenberg chain, unless stated otherwise.
Using a Jordan-Wigner transformation, this model can be mapped to the model of nearest-neighbor interacting hard-core bosons with an on-site disordered potential.
We consider the zero spin sector $\sum_i \langle S_i^z \rangle = 0$, which corresponds to half-filling in the particle picture.
The Anderson model of non-interacting particles in a disordered potential is retrieved for $\Delta = 0$.

The model (\ref{eq:hamiltonian}) was used to discuss the MBL transition and the subdiffusive behavior in the delocalized phase, see the references above. 
{Specifically, in the model \eqref{eq:hamiltonian}, an MBL regime, characterized by area-law entanglement, has been identified for strong disorder.
On the other hand, for weak disorder an ergodic regime has been reported, described by the eigenstate thermalization hypothesis (ETH) \cite{Deutsch1991a, Srednicki1994a} and exhibiting volume-law entanglement. }

\subsection{Time-dependent variational principle}

To study the model \eqref{eq:hamiltonian}, we employ a recently developed numerical method for describing the time evolution of 1D lattice systems, which is based on the Dirac-Frenkel time-dependent variational principle (TDVP) \cite{Dirac1930a} as applied to MPS \cite{Haegeman2011a, Haegeman2016a}. Contrary to traditional time-evolution algorithms, such as the time-dependent density matrix renormalization group (t-DMRG) \cite{Daley2004a,White2004a} or time-evolving block decimation (TEBD) \cite{Vidal2003a}, the TDVP does not rely on a Suzuki-Trotter decomposition of local Hamiltonian terms. Instead, the time-dependent wave function $\rpsi (t)$ is given by
\begin{equation}
 \frac{\dd \rpsi}{\dd t} = -\mathrm{i} \mathcal{P}_{\mathrm{MPS}} \mathcal{H} \rpsi,
\end{equation}
where $\mathcal{P}_{\mathrm{MPS}}$ projects the time-evolved wave function back onto the variational MPS manifold, with a dimension typically much smaller than the dimension of the complete Hilbert space $2^L$.

From the perspective of accuracy of MPS simulations, the worst-case scenario for a local Hamiltonian is realized when the von Neumann entanglement entropy for a bipartition of the chain into two parts grows linearly \cite{Calabrese2005a}, which is known as \emph{volume-law entanglement} and should be contrasted to the \emph{area-law entanglement} characteristic of localized systems \cite{Laflorencie2016a}.
It is well known that in such a worst-case scenario, the required bond dimension $\chi$ (which controls the dimension of the variational manifold) grows exponentially in time if it is required that the truncation error is kept below some finite value. This provides a fundamental limit to the maximum time reached using MPS simulations under the condition that $\langle \psi_\textrm{MPS} | \psi_\textrm{exact} \rangle \approx 1$ \cite{Schollwock2011a}.

Within the traditional DMRG framework, recent developments have extended the maximum possible times significantly \cite{Karrasch2013a}, although the fundamental limit remains.
On the other hand, under certain conditions it is expected \cite{Leviatan2017a, Kloss2017a, Berta2017a} that the TDVP can provide reasonable estimates for the transport properties despite a potentially large truncation error. {We stress, however, that this distinction between the TDVP and ``traditional'' MPS methods is only relevant relatively deep in the ergodic regime. In the MBL phase, the entanglement entropy grows logarithmically with time \cite{Bardarson2012,Serbyn2013b,Laflorencie2016a} and reaches the cutoff set by the bond dimension at times that are typically much longer than the simulation time. 
This renders the ``traditional'' MPS methods accurate in the MBL phase \cite{Khemani2016a,Pollmann2016}. Importantly, close to the MBL transition, entanglement growth is slow even on the ergodic side, and, in this critical regime, the TDVP is expected to be reliable in a large time window.} 
A key advantage of the method is that unlike t-DMRG and TEBD, it conserves the total energy by construction.

In the present work, we focus on the range of sufficiently strong disorder $W \ge 2$, which includes the whole MBL phase as well as a part of the delocalized phase. As we have verified (see detailed explanations {in section \ref{sec: numresultstdvp} below and in the Appendix}), the MPS approach indeed works reliably up to long times considered in our study not only in the MBL phase but also in the delocalized phase with $W \ge 2$ since the corresponding dynamics is very slow.

\begin{figure*}[!htb]
\includegraphics[width=\textwidth]{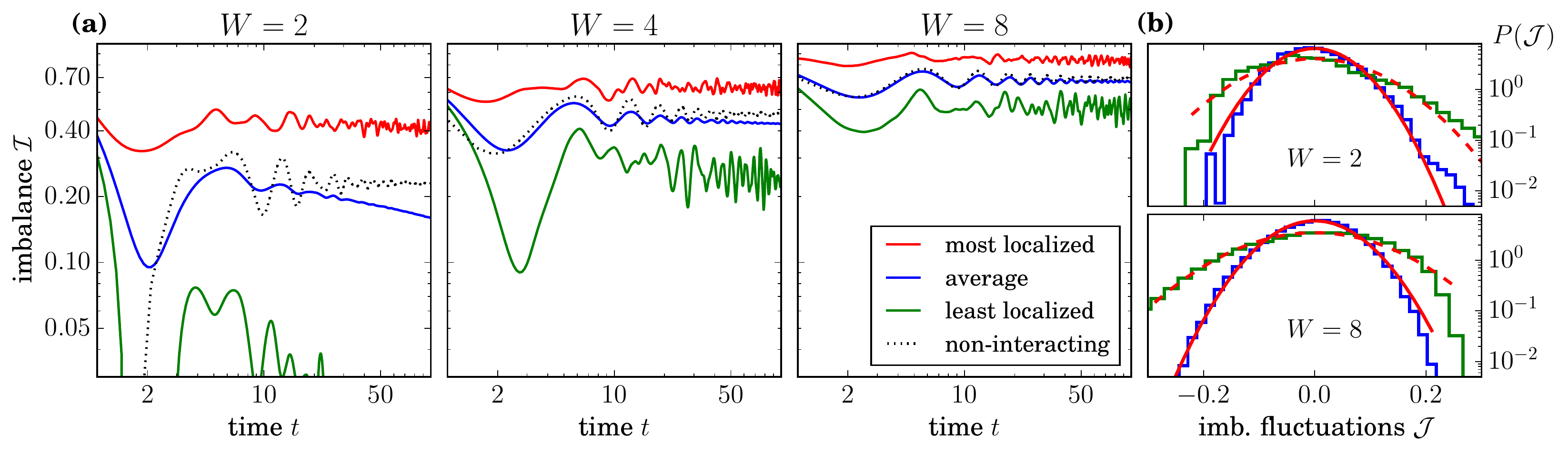}
 \caption{\textbf{(a)} Spin imbalance \eqref{eq:imba} as a function of time $t$ for a spin chain of length $L=100$ and various disorder strengths $W = 2, 4, 8$. The blue line shows the disorder average over $R = \mathcal{O}(10^3)$ realizations and the dotted black line shows the non-interacting case $\Delta = 0$ using $10000$ independent disorder realizations. The red (green) line shows the most (least) localized single realization. \textbf{(b)}  Probability distribution function of the fluctuations of the imbalance around the average $\mathcal{J}(t) \equiv \mathcal{I}(t) - \overline{\mathcal{I}(t)}$ for $W = 2$ (top), $W = 8$ (bottom), $L=50$ (green) and $L=100$ (blue) over the time interval $t \in [50, 100]$. The solid (dashed) red line shows a Gaussian fit for $L=100$ ($L=50$).}
\label{fig:denstime}
\end{figure*}

\begin{figure}[!htb]
 \includegraphics[width=\columnwidth]{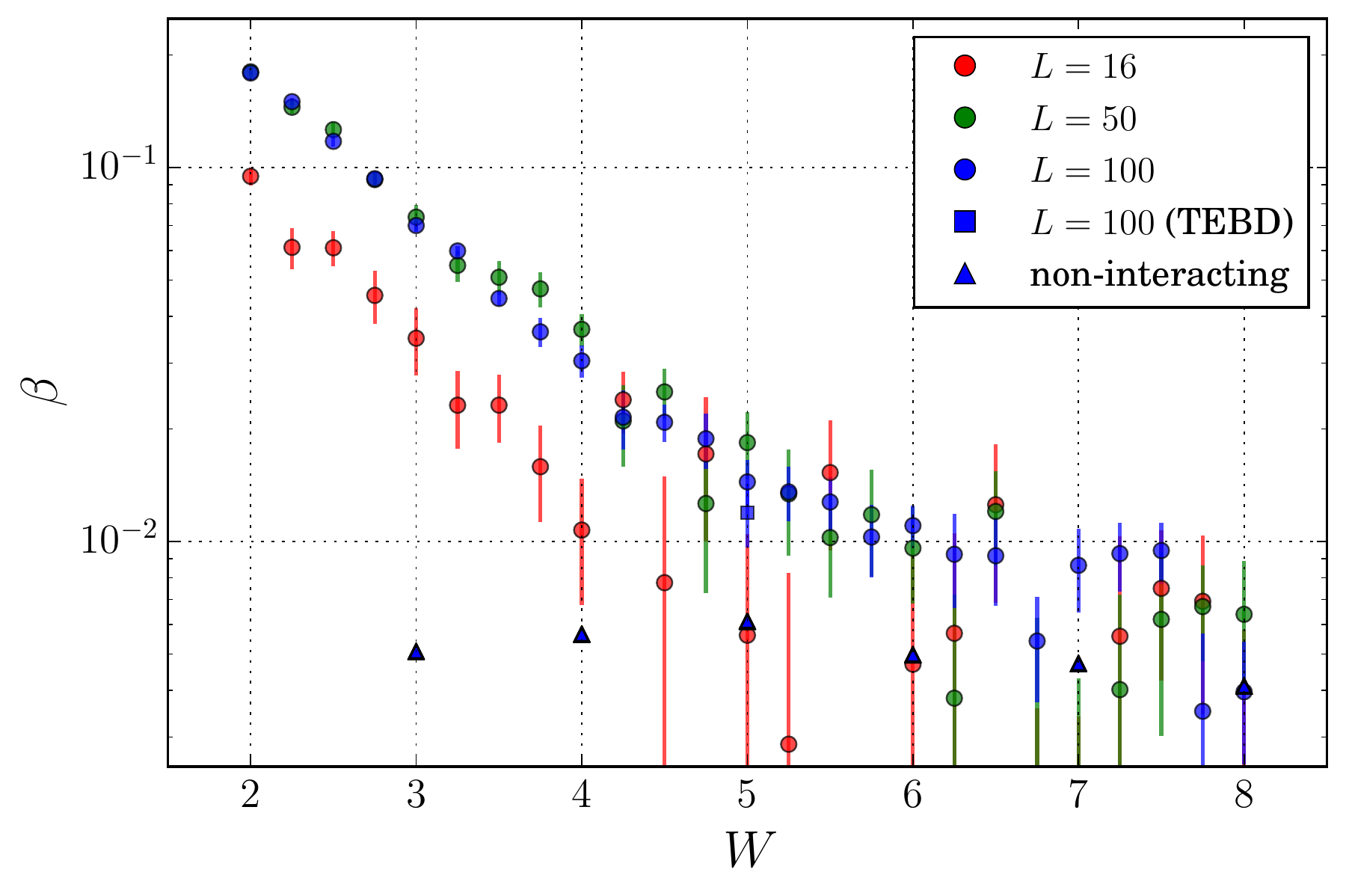}
 \caption{Power-law exponent $\beta$ corresponding to the decay of the imbalance 
 $\mathcal{I} \propto t^{-\beta}$ over the window $t \in [50, 100]$, for various system sizes $L$. For the cases $L=50$ and $100$,
 the 10 sites near each edge of the chain were not considered. The square symbol shows the result for an independent implementation of the time-evolving block decimation (TEBD) algorithm \cite{Jaschke2018a}. Triangles indicate a weak finite-time decay of $\mathcal{I}_\mathrm{A}$ for the non-interacting case $\Delta = 0$, $L=100$. Error bars are $1\sigma$-intervals based on a bootstrapping procedure.}
 \label{fig:beta}
\end{figure}

\section{Numerical results (TDVP)}
\label{sec: numresultstdvp}

In this section, we will detail results obtained using the direct analysis of the TDVP data for the quench 
dynamics in disordered Heisenberg chains \eqref{eq:hamiltonian} of lengths $L=16,\ 50$ and $100$ with disorder strength ranging from $W=2$ to $W=8$.

\subsection{Imbalance}

We follow the time dynamics of an initial unentangled (product) N\'eel state $\rpsi = \{ \uparrow, \downarrow, \ldots \}$, where we consider many different realizations $R \gg 1$ of disorder.
Whether the system is in the delocalized or localized phase can be quantified using the \emph{imbalance}:
\begin{equation}
 \mathcal{I}(t) = \frac{1}{L} \sum_{i=1}^L (-1)^i \langle S^z_i (t) \rangle, \label{eq:imba}
\end{equation}
a quantity that measures how much of the initial antiferromagnetic order remains at time $t$ \cite{Luitz2017a}. 
By definition, $\mathcal{I}(0) = 1$, and for a fully ergodic, thermalized state, $\mathcal{I} (t) \to 0$ at long times on average, which means that the system completely loses memory of the initial antiferromagnetic order.
The disorder average is denoted as $\overline{\mathcal{I}(t)} \equiv (1/R) \sum_{r} \mathcal{I}_r(t) $, where $r = \{1, \ldots, R\}$ labels the different realizations.

The imbalance is appealing from several perspectives.  It is a global characteristic of the system which 
signifies its degree of localization, but at the same time it is computed from purely local quantities (average spins). In view of this locality, the imbalance can be readily measured in an experimental setting using cold atoms \cite{Schreiber2015a}. We can also compare directly the localization properties of an interacting system  to those of non-interacting Anderson insulators. 
The non-interacting value $\mathcal{I}_\mathrm{A}(t)$ is obtained by computing the time evolution of the N\'eel state through the exact diagonalization of \eqref{eq:hamiltonian} with $\Delta = 0$  (cf.\ Ref.\ \cite{Berkelbach2010a}).
In Fig.\ \ref{fig:denstime}, we show the dynamics of imbalance for both the interacting and non-interacting cases using the parameters $L = 100$, $W = \{2, 4, 8\}$.

For a thermalized system, a power-law decay $\mathcal{I}(t) \propto t^{-\beta}$ is expected \cite{Luitz2017a}.  In fact, within the Boltzmann equation, one would get an exponential decay with time, since the imbalance corresponds to a mode with a large wave vector $q$  (namely, $q = \pi$) describing, in contrast to the total spin, a non-conserved quantity. However, taking into account the coupling of this mode to the low-$q$ diffusive (or subdiffusive) mode associated with the spin density spreading suggests a power-law decay (by analogy with the long-time tails related to the return probability that are found for other observables). Indeed, previous numerical results indicated that the exponent $\beta$ of the imbalance decay is qualitatively similar to the subdiffusion exponent found from the return probability, the mean-square displacement of a spin excitation, and the low-frequency dependence of the conductivity (see Ref.~\cite{Luitz2017a} and references therein).

In Fig.\ \ref{fig:beta}, we show the exponent $\beta$ computed numerically using a least-squares fitting algorithm for various values of $W$ and $L$ from the imbalance at sufficiently long times $t \in [50, 100]$.
We choose $W = 2$ as the weakest disorder to be considered. This is based on the requirement that the value 
$\chi = 64$ of the bond dimension used in our computations is sufficient to ensure that the value of $\beta$ is insensitive to a further increase of $\chi$ (see Appendix \ref{sec:appendix}).
As a further check, we compute $\beta$ using a completely independent implementation of the time-evolving block decimation (TEBD) algorithm for $W = 5$, $L = 100$, $\chi = 64$, finding agreement within error bars.
In addition, for a few values of $W \geq 4$ and for the system size $L=100$ we follow the time evolution until $t = 300$ using $\chi = 32$ and find a good agreement with $\beta$ obtained in the window $t \in [50,100]$ (not shown), as well as checking that for bond dimension up to $\chi = 96$, $\beta$ is insensitive to increasing the time window up to $t = 200$ for $L = 50$ and $W = 5$ (see Appendix \ref{sec:appendix}).
The number of different disorder realizations for each choice of $W$ and $L$ is typically of the order of $500 - 1000$.

\begin{figure*}[!htb]
 \includegraphics[width=2.0\columnwidth]{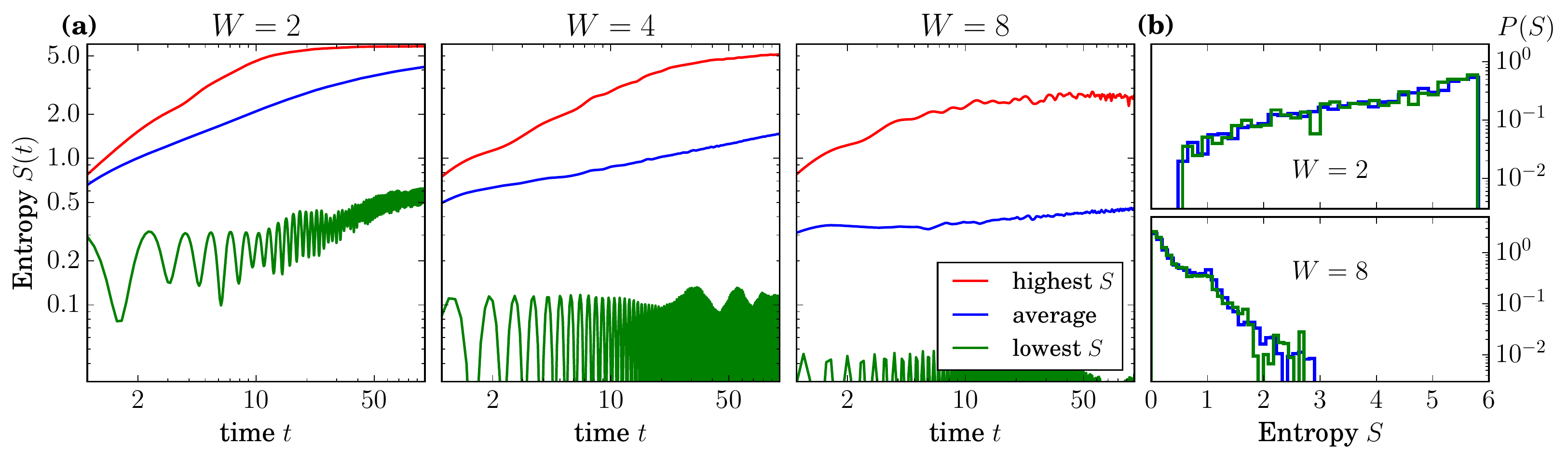}
\caption{\textbf{(a)} 
The von Neumann entropy of entanglement $S$ as a function of time
for the same parameters as in Fig.\ \ref{fig:denstime}. \textbf{(b)} Distributions of $S$ over the interval $t \in [95,100]$. The blue (green) lines show the result for $L=100$ ($L=50$).}
 \label{fig:enttime}
\end{figure*}

For small systems that can be treated by exact diagonalization (such as $L = 16$), we find a good agreement with previous results~\cite{Luitz2016a}.  Specifically, the value of $\beta$, which continuously decreases with increasing $W$, vanishes within error bars at $W \approx 4$.  However, an  increase of the  system size from $L=16$ to $L=50$ leads to a substantial increase of $\beta$, consistent with a trend observed for relatively small systems in Ref.~\cite{Bera2017a}. As a result, the value of disorder for which $\beta$ vanishes also increases
\footnote{It should be noted that for small $\beta = \mathcal{O}(10^{-2})$, we can no longer distinguish a power-law decay over our limited time window from, say, a logarithmic correction.
Nevertheless, this procedure yields a sensitive probe for the decay of $\mathcal{I}(t)$ in the sense that \textit{any} statistically significant monotonous decay will result in a non-zero $\beta$.}.
{This is also consistent with the findings in Ref.~\cite{Pietracaprina2017a}, which suggest that exact diagonalization approaches are subject to strong finite-size and boundary corrections.}
On the other hand, the results for $\beta$ obtained for $L=50$ and $L=100$ agree within error bars.  
We have further diminished finite-size effects by excluding the 10 sites on each far end of the chain for the cases $L=50,100$ for the computation of $\mathcal{I}$, so that we only consider the bulk of the system. 

We thus conclude that, after the initial increase for $L \lesssim 50$, the exponent $\beta$ saturates, i.e., essentially reaches its thermodynamic-limit ($L \rightarrow \infty$) value. This implies two important conclusions. First, the slow, subdiffusive transport appears to be a genuine property  of a long chain on the ergodic side of the MBL transition and not just a finite-size effect. Second, the estimated value of the critical disorder is  $W_c \gtrsim 5$, i.e., considerably larger than the values suggested by the exact-diagonalization analysis of small systems. {More precisely, an estimated lower bound is determined by considering the lower end of our error estimate with two standard deviations, plus a small systematic error shown as triangle symbols in Fig.~\ref{fig:beta} \footnote{The imbalance of a non-interacting system $\mathcal{I}_\mathrm{A}(t)$, while asymptotically approaching a positive constant, shows a weak decay with time. We have checked that fitting of this decay by a power law in the considered time interval yields fictitious values of $\beta$ in the range $\beta \approx 0.004-0.006$. This implies a lower bound on the true exponent $\beta$ of a many-body delocalized system that we can distinguish from zero.}, which yields a lower bound of $W_c \approx 5.5$--$5.75$ (see Appendix). Requiring a higher confidence of $4\sigma$-intervals yields $W_c \approx 5$.} For weaker disorder, we can confidently conclude the system is ergodic.

In order to understand how representative  the average value of the imbalance is,  we have also studied its fluctuations, $\mathcal{J}(t) \equiv \mathcal{I}(t) - \overline{\mathcal{I}(t)}$. The root-mean-square amplitude of the fluctuations of the imbalance, $\sqrt{ \overline{ \mathcal{J}^2(t)}}$, has been found to be very slowly varying with time for long times. This allows us to consider the probability density function $P[\mathcal{J}(t)]$, where $t \in [50, 100]$.
In this window, $\mathcal{J}(t)$ is, to an excellent approximation, Gaussian-distributed (see Fig.\ \ref{fig:denstime}b), with a standard deviation $\sigma$ roughly proportional to $1/\sqrt{L}$. This implies self-averaging of the imbalance. A similar behavior is found for a non-interacting, Anderson-localized system.
The wide distributions for smaller system sizes complicate the accurate determination of $\beta$ around the transition, so that considering larger system sizes is very beneficial also from this point of view.

\subsection{Entanglement and Schmidt gap}

The entropy of entanglement characterizes the spread of correlations in the system.
On the ergodic side, power laws have been predicted for the growth of entanglement close to the transition \cite{Luitz2016a}: $S(t) \propto t^{\beta_S}$.
Another quantity of interest is the so-called Schmidt gap $\Lambda$, which is defined as the difference between the two largest values of the entanglement spectrum (see Ref.\ \cite{Laflorencie2016a} and references therein) for a bipartition of the system into subsystems $A$ and $B$: $\Lambda = \lambda_1 - \lambda_2$. 
Here the entanglement eigenvalues $\lambda_i\geq 0$ \footnote{In numerical language, the entanglement eigenvalues are also the squares of the Schmidt numbers, obtained easily in the MPS representation through singular value decompositions.} are the eigenvalues of the effective entanglement Hamiltonian $H_e$ defined through the reduced density matrix of the system after tracing out the degrees of freedom of the subsystem $B$: $H_e = -\ln \text{Tr}_B \rho$ \cite{Li2008a}.

\begin{figure*}[!htb]
 \includegraphics[width=2.0\columnwidth]{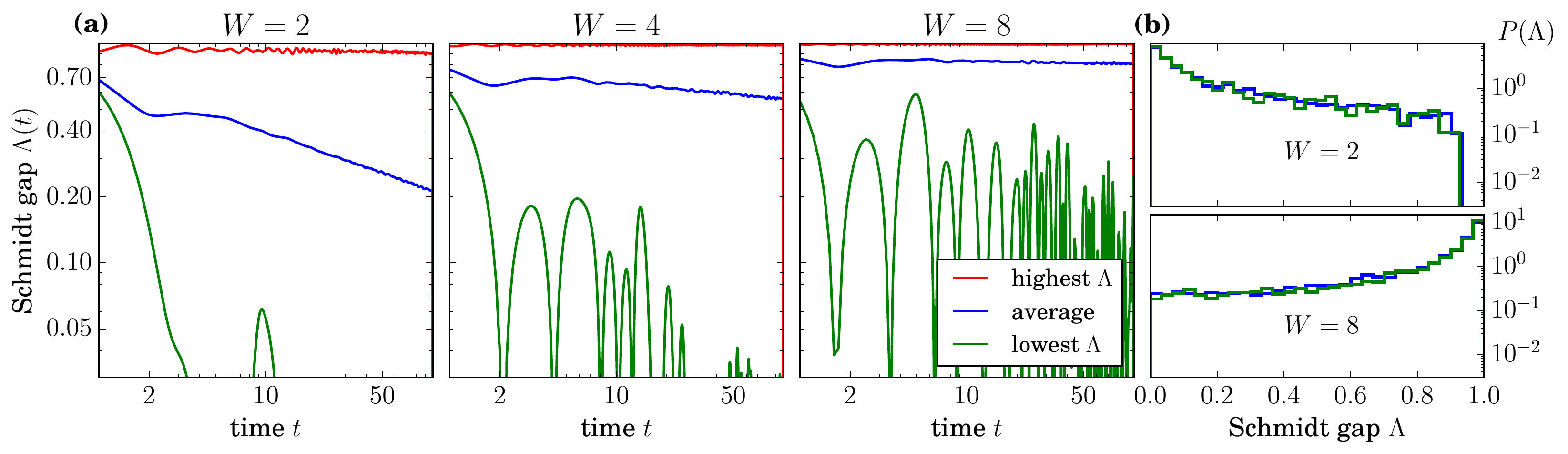}
\caption{\textbf{(a)} The Schmidt gap $\Lambda$ as a function of time
for the same parameters as in Fig.\ \ref{fig:denstime}. \textbf{(b)} Distributions of $\Lambda$ over the interval $t \in [95,100]$. The blue (green) lines show the result for $L=100$ ($L=50$).}
 \label{fig:schtime}
\end{figure*}

\begin{figure}[!htb]
 \includegraphics[width=\columnwidth]{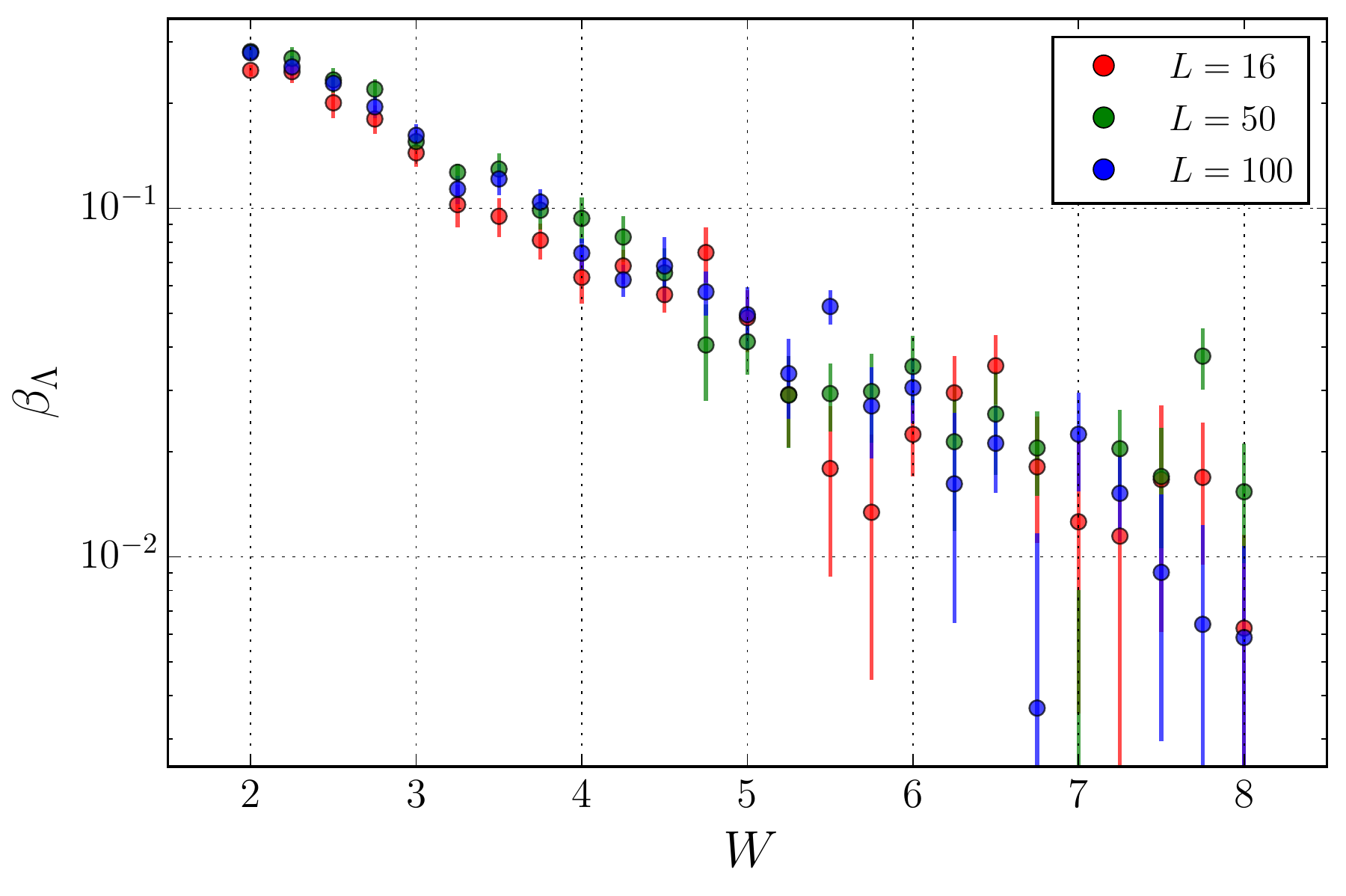}
 \caption{Power-law exponent $\beta_\Lambda$ corresponding to the decay of the disorder-averaged Schmidt gap 
 $\Lambda \propto t^{-\beta_\Lambda}$ over the window $t \in [50, 100]$, for various system sizes $L$. Error bars are $1\sigma$-intervals based on a bootstrapping procedure.}
 \label{fig:schmidt}
\end{figure}

To explain the connection between the Schmidt gap and the more frequently used 
von Neumann entropy of entanglement, we recall that the latter can be expressed in terms of the entanglement eigenvalues $\lambda_i$ as follows :
\begin{equation}
 S = -\sum_{i} \lambda_{i} \log_2 \lambda_i  \simeq  -\sum_{i=1}^\chi \lambda_{i} \log_2 \lambda_i. \label{eq:entropy}
\end{equation}
Here the eigenvalues $\lambda_i$ are ordered in a descending way by convention. The approximate equality holds as long as the entanglement remains relatively low. For our problem, we find that the effect of the cutoff by the bond dimension $\chi$ is negligible for $W \gtrsim 4$ from the perspective of $S$ (see Fig.\ \ref{fig:enttime}).


\begin{figure*}[t]
\includegraphics[width=1.9\columnwidth]{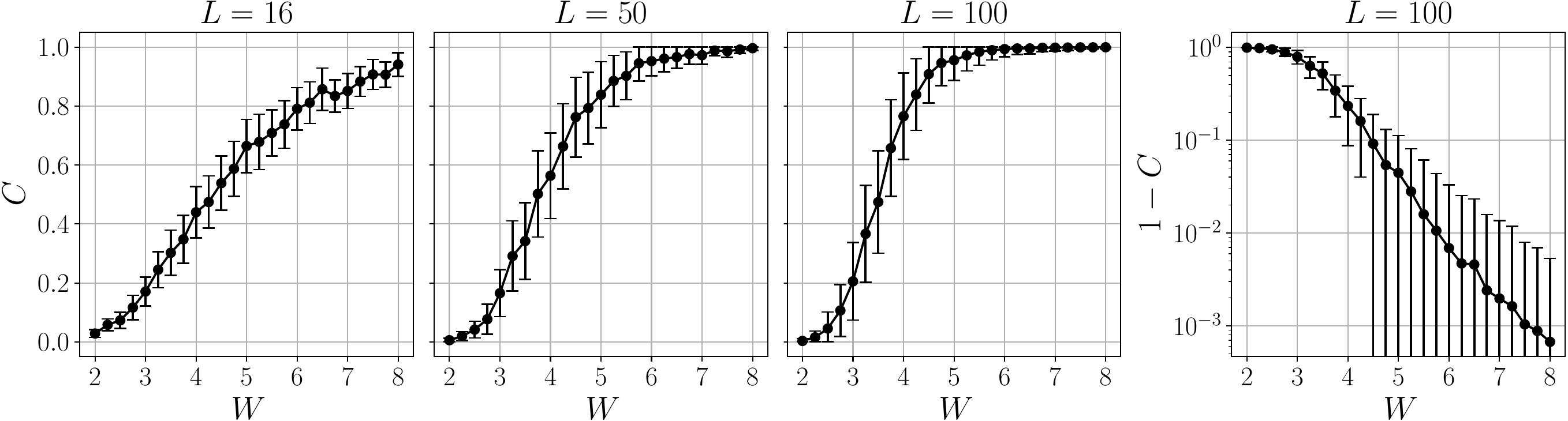}
\caption{
Results of the supervised machine learning algorithm. Shown is the average confidence $C$ with which time-series (spin densities $\langle S^z \rangle$ as a function of time) taken from TDVP for a given disorder magnitude are classified as belonging to the localized phase, where we labeled $W = 8$ as localized and $W = 2$ as delocalized. A plateau emerges at large $W$ for $L=100$, indicating a transition to the MBL regime. The rightmost panel shows $1-C$ on a semi-log scale. Error bars {indicate the standard deviation of the distribution of $C$ and} are $1\sigma$-intervals based on $100$ independent training sessions (these should be understood as being cut off at $C = 0$ and $C = 1$).
}
\label{fig:MLresults}
\end{figure*}


Contrary to the entanglement entropy, the evaluation of the Schmidt gap requires knowledge of only the first two entanglement eigenvalues and hence is less sensitive to the value of the bond dimension.
A further appealing property of this quantity is that, for a thermalized system, $\Lambda(t) \rightarrow 0$ for $t\to \infty$, while for a localized system it is expected that the Schmidt gap remains finite in the long-time limit, $\Lambda(t) \rightarrow \mathcal{O}(1)$, or at most decays logarithmically. In this sense, the $t\to \infty$ behavior of the Schmidt gap allows one to distinguish between the localized and delocalized phases in the same way as for the imbalance.

The entanglement entropy and Schmidt gap for \mbox{$L=100$} and representative choices of $W$ are depicted in Figs.\ \ref{fig:enttime} and \ref{fig:schtime}, respectively, which also show the distributions of these quantities close to the final time $t=100$ considered here. The shape of the distributions found for the entropy of entanglement is in qualitative agreement with a detailed study of such distributions for eigenstates of small systems using exact diagonalization \cite{Luitz2016b}. That these distributions are converged with respect to system size reflects the slow spread of entanglement close to the transition (even on the ergodic side). {The entanglement itself is thus not suitable for determining the location of the transition, since it is slowly growing on both sides of it. The Schmidt gap, however, is more promising in this regard.}

We have determined the Schmidt gap by using a bipartition in the middle of the chain. Interestingly, we find that the averaged Schmidt gap shows in the delocalized phase a power-law decay similar to that as $\mathcal{I}(t)$.  The corresponding power-law exponent $\beta_\Lambda$ is shown in Fig.\ \ref{fig:schmidt}.
The results demonstrate a striking qualitative similarity to those for the imbalance exponent $\beta$, Fig.\ \ref{fig:beta}.  However, the found large-$L$ values of $\beta_\Lambda$ are somewhat above those for $\beta$. The corresponding estimate for the critical disorder of the MBL transition obtained as a point where the Schmidt gap exponent $\beta_\Lambda$ vanishes within error bars is $W_c \simeq 6$, i.e., even larger than $W_c \simeq 5$ found from results for the imbalance. {We note that a recent study of the Schmidt gap in relatively short chains up to $L=20$ \cite{Gray2018a} also found evidence for an increase of the critical disorder compared to other methods. This is consistent with Fig.\ \ref{fig:schmidt}, where we show that $\beta_\Lambda$ is less sensitive to system size than $\beta$.}
For the entropy of entanglement $S$, however, we clearly see the effect of the cutoff at $S = 6$ imposed by the bond dimension $\chi = 64$ (see Fig.\ \ref{fig:schtime}a, left panel), where the power-law behavior is disturbed around $t\approx 30$.
Considering that the transport properties have nonetheless converged with $\chi$, this indicates that the TDVP can indeed provide reasonable results even if there are significant cutoff effects in terms of the entropy.
Moreover, the Schmidt gap, which still shows a clean power-law behavior at $W=2$, appears to be less sensitive to this numerical cutoff. {This can be understood as a consequence of the fact that the Schmidt gap is not \emph{directly} affected by the cutoff of the entanglement spectrum [see Eq.~\eqref{eq:entropy}], whereas the entropy is. Note, however, that the choice of bond dimension and the minimum $W = 2$ is based on transport properties (see Appendix).}

\section{Machine learning}

To further corroborate our analysis, we apply methods from machine learning \cite{Bishop2006a}, 
which has emerged recently as a powerful tool to analyze localization phenomena \cite{Schindler2017a, Mano2017a, Venderley2017a, vanNieuwenburg2018a, Hsu2018a, Zhang2018a}, to our data obtained using the TDVP.
We use two algorithms: a partially supervised approach that has previously been employed in Ref.~\cite{Schindler2017a}, and a fully unsupervised method based on the ``learning by confusion'' scheme introduced in Ref.~\cite{Nieuwenburg:2017qy}.
The combination of traditional numerical analysis and machine learning is mutually reinforcing: understanding localization though machine learning amounts to learning machine learning through localization.

\subsection{Supervised classification algorithm}
\label{sec: supervised ML}

For the supervised learning approach, we train a feed-forward neural network to distinguish data at two extremes of our data set: $W = 2$ (delocalized) and $W = 8$ (presumed to be localized). We choose a single hidden layer network with a $\mathrm{ReLU}$ activation function for the hidden layer (of size $\sim 10$) and a $\mathrm{Softmax}$ activation function for the output layer (of size $2$, corresponding to the two classes we are distinguishing). As input data, we use time-series $\langle S^z_i (t) \rangle$, $i = 1 \dots L$, evaluated at equidistant time-steps $t \in [50, 51, \dots, 100]$ taken from the full TDVP time evolution at $W = 2$ and $W = 8$. We choose a simple cross entropy error function with $\ell_2$ regularization, cf.~Ref.~\cite{Schindler2017a}.

After convergence of the training set error, we apply the trained network on a test set containing data at $W = 2$ and $W = 8$, finding a classification accuracy of larger than $99\%$. We subsequently apply the trained network to TDVP time-series at intermediate disorder strengths $W = 2.25, 2.5, \dots, 7.75$, to determine the average confidence $C$ with which data corresponding to a given disorder strength is classified as either belonging to the delocalized ($W = 2$) or localized ($W = 8$) class of time-series we trained with. Successful training implies that $C=0$ for $W = 2$ and $C=1$ for $W = 8$, while $C \in [0,1]$ at intermediate disorder strengths quantifies how similar the time evolution is to the extreme values $W = 2$ and $W = 8$.
The results are shown in Fig.~\ref{fig:MLresults}.

This approach clearly indicates the formation of a plateau for $L=100$, suggesting the presence of a many-body localized phase over a range of disorder strengths, but its thermal counterpart gives way to an extended crossover region. In contrast, no plateau or a less pronounced one is present in the data for system sizes $L=16$ and $L=50$, respectively, showing the importance of considering larger systems.

These results can be interpreted in analogy with the analysis of the power-law exponents $\beta$ and $\beta_\Lambda$: over a range of disorder strengths, the imbalance does not decay and the spin densities ``look similar'' to the case $W = 8$, down to the critical disorder. However, the algorithm can still pick up differences between a large value of the power law and a small one, leading to a non-zero $C$ even in the delocalized regime, so that the behavior of $\beta$ and $1-C$ is qualitatively similar (see Fig.~\ref{fig:MLresults}, rightmost panel). The plateau, within error bars, starts at $W \gtrsim 4.5$ \footnote{The start of the plateau can be associated with the value of $W$ where the lower bound of the error bar in the right panel of Fig.~\ref{fig:MLresults} touches the upper bound of the error bar for the training at $W=8$.}. Hence, one should not expect the shape of the curves to approach a step-like function in the limit $L \rightarrow \infty$ (the sharpening with system size can be associated with the narrowing distributions $\mathcal{J}$, see Fig.\ \ref{fig:denstime}b), in contrast to level statistics \cite{Luitz2015a}. A precise determination of the critical disorder strength, relying on knowledge of the extent of the ``plateau'' in the limit $L \rightarrow \infty$, is therefore difficult to obtain using this approach, and placing the transition for instance at the midpoint $C = 1/2$ leads to an underestimate for the critical disorder \cite{Schindler2017a}. Nonetheless, in the region where $\beta$ vanishes within error bars, the supervised learning approach is consistent with the analysis of the decay of the imbalance in the sense that in this region (belonging to the MBL plateau, as defined above) $C \gtrsim 0.99$ for $L=100$. In the next section, we will detail an unsupervised method that is able to determine $W_c$ more precisely.


\begin{figure*}[t]
\includegraphics[width=0.95\textwidth]{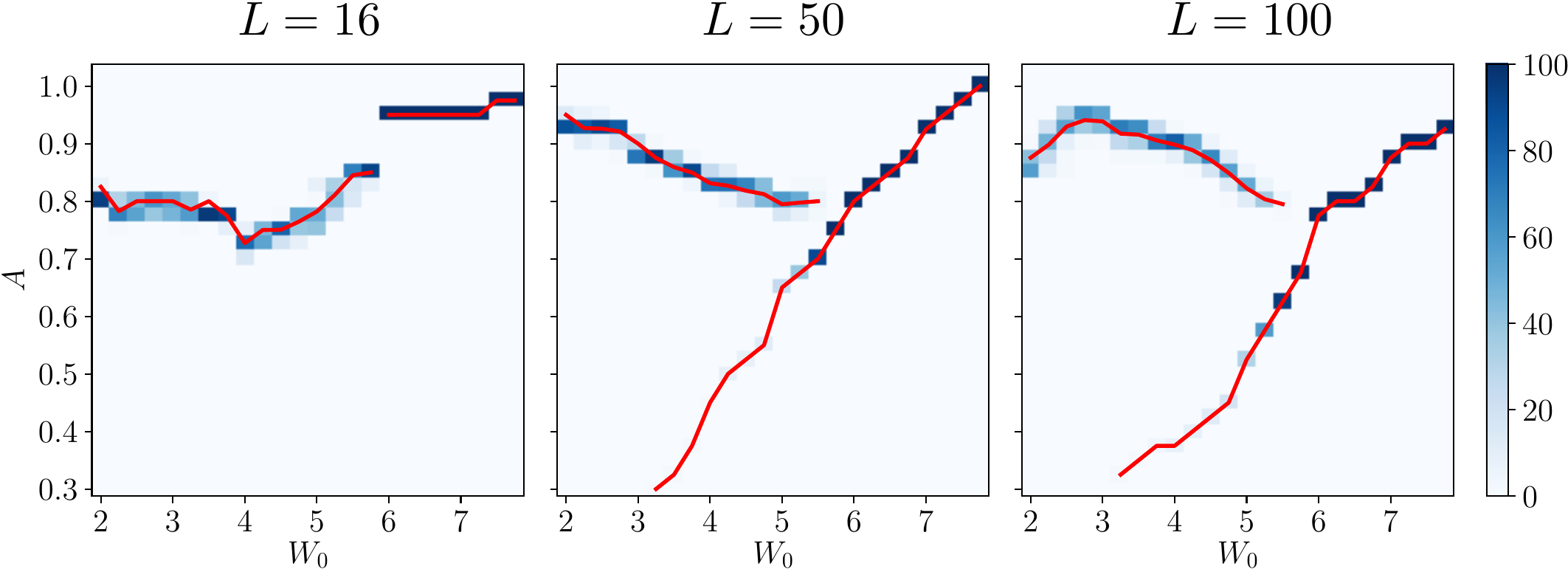}
\caption{
Results of the unsupervised machine learning algorithm, based on ``learning by confusion'' for input data with $L=16$, $L=50$, and $L=100$. 
Shown is the distribution of the accuracy $\langle A(W_0)\rangle$ over 100 training instances with different initial conditions for each $W_0$. For system sizes $L=50$ and $L=100$, the distribution is bimodal in the interval $3.25<W_0<5.5$ with the two maxima tracing out the red lines as a function of $W_0$. 
}
\label{fig:MLresults-2}
\end{figure*}


\subsection{Unsupervised confusion algorithm}
In Ref.~\cite{Nieuwenburg:2017qy} a scheme called ``learning by confusion'' for the unsupervised detection of phase transitions for data ordered along a one-dimensional parameter space (here: the strength of disorder $W\in [W_{\mathrm{min}},W_{\mathrm{max}}]$) has been proposed. For this, an arbitrary parameter $W_0$ is fixed and a feed-forward neural network is trained (in a supervised fashion) to distinguish data with $W<W_0$ from data with $W>W_0$. The accuracy $\langle A(W_0)\rangle$ of classification that the trained network achieves when applied to a test set is evaluated. The process is repeated for different choices of $W_0$. The resulting function $\langle A(W_0)\rangle$ has two global maxima at $W_0=W_{\mathrm{min}}$ and $W_0=W_{\mathrm{max}}$, where the accuracy  is trivially 1 because all the data can be classified as belonging to one phase. In Ref.~\cite{Nieuwenburg:2017qy} it was observed that an additional local maximum of $\langle A(W_0)\rangle$ occurs when $W_0$ equals the location of a phase transition $W_\mathrm{c}$, as it would be easiest for the network to classify the data into two sets for this choice of separation. Thus, in presence of a single phase transition as a function of $W_0$, the curve $\langle A(W_0)\rangle$ is expected to take a W-shape.

We apply this algorithm, using the same neural network architecture and the same type of input data as in the supervised case, while dividing the training data by $W_0=2.25,2.5,\dots,8$. Each time we determine the average accuracy $\langle A(W_0)\rangle$ as the percentage of
correctly classified sets $\langle S^z_i (t) \rangle$ (with respect to the division $W_0$) in a test set (which does not coincide with the training data).
We observe that the results depend strongly on the initial conditions for the training, i.e., on the random initial choice of weights and biases for the network. Even for input data from the largest system sizes $L=100$, no consistent W-shape emerges in $\langle A(W_0)\rangle$. This is due to substantial fluctuations of $\langle A(W_0)\rangle$ for fixed $W_0$ between training runs, in particular around the putative transition region. 

The fluctuations of $\langle A(W_0)\rangle$ resulting from the initial conditions of the training are, however, not random. We claim that they carry information that can be used to locate the transition. We observe that over a large region of $W_0$, they follow a bimodal distribution as shown in Fig.~\ref{fig:MLresults-2}. We interpret the two branches as instances of trained networks that identify localized and ergodic features, respectively
\footnote{It is interesting to note that the ``MBL branch'' for large chains ($L=50$ and $100$) starts at $W\simeq 3-4$, which is close to the estimate for the critical disorder that was previously obtained from the exact diagonalization in relatively small chains.}. The transition should then be identified as the position where both types of networks occur with the same probability, which happens around $W_0\approx 5$. Viewing the ETH-MBL transition as a crossover (at least from the perspective of the finite-size and finite-time data that serves as input of the neural network), this analysis puts $W_0\approx 5.5$--$6$ as an upper bound to the crossover region. Above this value the distribution of  $\langle A(W_0)\rangle$ becomes unimodal.

\section{Summary and Outlook}

The quench dynamics of large disordered spin chains in the Heisenberg model has been investigated by means of the time-dependent variational principle for matrix product states. We have studied the long-time behavior of the imbalance and found  a regime, occurring in a broad range of parameters of the system, with slow, yet finite transport. We find that the average, typical and median imbalance all lead to the same power-law exponent $\beta$. Our results imply that the ergodic regime extends (at least) up to disorder $W_c \simeq 5$.  We observe a substantial shift of the exponent $\beta$ (and, as a result, of an estimate for the MBL transition point $W_c$) when we go from relatively small systems (that can be exactly diagonalized) to large systems with $L = 50$ and $100$.  On the other hand, we do not see any significant difference between the results for $L=50$ and $L=100$, which indicates saturation of the exponent $\beta$. This favors the conclusion that the subdiffusive behavior is a true long-time asymptotic behavior. These findings have been supported by the results for the dynamics of the Schmidt gap in the entanglement spectrum, which shows a very similar behavior. 

{Our analysis, demonstrating ergodic behavior in large spin chains for disorder up to $W_c \simeq 5$,
substantially shifts the commonly quoted estimate for the MBL transition ($W_c\simeq 3$--$4$) that was previously obtained (see, e.g., Ref. \cite{Luitz2015a}) by exact diagonalization of XXZ chains of length up to $L\sim 20$. This advance is not only quantitative, but it also implies an important qualitative statement concerning the nature of the MBL transition. Indeed, enhanced ergodicity in larger systems supports the existence of ``non-local'' (involving states distant in real space) delocalizing processes that are not typically present in small systems.}

{On the methodological side, our results demonstrate the reliability of the TDVP approach for studying the
XXZ model at not too weak disorder ($W\geq 2$ in our case), in agreement with prior findings \cite{Kloss2017a} concerning the applicability of TDVP to this model. This range of disorder strength includes both sides of the MBL transition. 
In this range of $W$, the system is characterized by a slow growth of entanglement with time, which has allowed us to controllably explore the quench dynamics in large chains (with lengths inaccessible by exact-diagonalization methods) within the time window sufficient to infer the behavior of the system in the thermodynamic limit. This opens new opportunities for computational studies of correlated disordered models by means of MPS-based approaches.}

We have complemented a conventional analysis of the data (with fitting the average values to power laws) by a machine-learning analysis of the whole time dependences of individual spins in various realizations of disorder. We have chosen two approaches, a supervised and a novel unsupervised algorithm. The latter is based on learning by confusion, but also exploits the stochastic nature of the learning process. This approach provides a more accurate way to determine the position of the transition, and could be useful for wider applications in determining phase transitions.
The results from both the supervised and unsupervised machine-learning methods support our conclusions concerning the bound for the critical disorder strength $W_c \gtrsim 5$.
Thus, we can reinforce machine learning by combining it with a ``traditional'' analysis of numerical data, verifying that machine-learning tools can be reliably applied.

Our results, showing a very slow transport on the ergodic side of the MBL transition ($W < W_c$) support the expectation that the system looks essentially localized at the transition point ($W= W_c$) \cite{Pal2010a}.  From this point of view, the MBL transition has much in common with the Anderson transition on random regular graphs (RRG)~\cite{Tikhonov16}. The latter problem is viewed as a toy model of the MBL transition, even though this connection is less precise for short-range interaction models than for those with power-law interactions~\cite{Tikhonov18}. A slow dynamics in the RRG model was recently studied numerically in Ref.~\cite{Biroli2017a}. From the perspective of the machine-learning approaches applied in this work, the notion that the transition point is localized in character manifests itself in that it is located near the edge of the ``plateau'' characterizing the MBL phase, rather than at the midpoint between the two regimes. In terms of the confusion-based algorithm, the transition is located at the crossing between ``delocalized'' and ``localized'' branches.

A slow dynamics near the transition and a localized character of the critical point are also qualitatively consistent with the avalanche mechanism of the transition developed recently in { Refs.~\cite{DeRoeck2016a,Thiery2017a,Thiery2017b, Goremykina2018a} }. This scenario predicts an extended weakly delocalized regime above the ``nominal'' threshold $W_c^{(0)}$ and implies that previous numerical results might be tainted by finite-size effects. Specifically, within this scenario, static ``ergodic spots'' (spatial regions with anomalously weak disorder) embedded in the nominally localized phase \cite{Gopalakrishnan2015a} thermalize the rest of the system below the true MBL transition at $W_c>W_c^{(0)}$. In small systems, such spots are typically not found
and the chains appear to be localized for $W>W_c^{(0)}$; however, with increasing $L$ the probability of finding ergodic spots increases and the true transition at $W_c$ becomes well resolved. {This is in line with the aforementioned importance of non-locality.}

Indeed, we find that considering larger systems leads to a substantial increase in delocalization, in agreement with the avalanche scenario. Moreover, the unsupervised machine-learning analysis, showing the coexistence of localizing and delocalizing realizations of disorder in the range $3.5\lesssim W\lesssim 5.5$ (see Fig.~\ref{fig:MLresults-2}), may also be interpreted in terms of the avalanche-induced delocalization. 
Further, consistent with the prediction of Thiery \textit{et al.}~\cite{Thiery2017a,Thiery2017b}, the system ``looks localized'' near the transition, as evidenced by the strong similarity between the spin dynamics for the non-interacting case $\Delta = 0$ and the interacting case close to our lower bound for the MBL transition.
However, at the present stage, we cannot explicitly confirm (or falsify) this mechanism, as this would require a specific analysis of observables that would have a distinct behavior within this mechanism.
Another possibility consistent with our results is the existence of a glass-like crossover regime \cite{Biroli2017a}.

Future work inspired by our results could focus on different energy densities in addition to just the center of the band that has been considered here, in order to map the transition line in the energy-disorder plane on the basis of data for large systems.  Further, it is also interesting and instructive to analyze the phase diagram of long XXZ chains in the interaction-disorder plane.
Clearly, a better analytical understanding of the slow dynamics near the MBL transition would be very important. In this context, it is interesting to note that a similar slow transport near the transition (or, more accurately, apparent transition) is also found for quasiperiodic systems \cite{Schreiber2015a} as well as in two-dimensional disordered systems \cite{Bordia2017a}, where the influence of rare bottlenecks is expected to be negligible.  It remains to be seen whether this slow transport has a common origin in all these situations. 

A so-called ``dreaming'' protocol within the neural-network framework, which generates the configurations that are representative for given phases, could be very useful
for developing the analytical theories of the subdiffusive phase. Furthermore, this type of machine-learning approach can be envisaged to simulate the dynamics at much longer times (and, perhaps, in much larger systems), which is currently inaccessible by the most advanced numerical tools. 
Finally, a hotly debated question is the very existence of the true ($L \to \infty$) MBL transition in various models. We hope that our work will pave the way for numerical  investigations of the physics associated with the MBL transition on large systems, which is of obvious importance for shedding light on this issue.


\begin{figure}[!htb]
\includegraphics[width=.95\columnwidth]{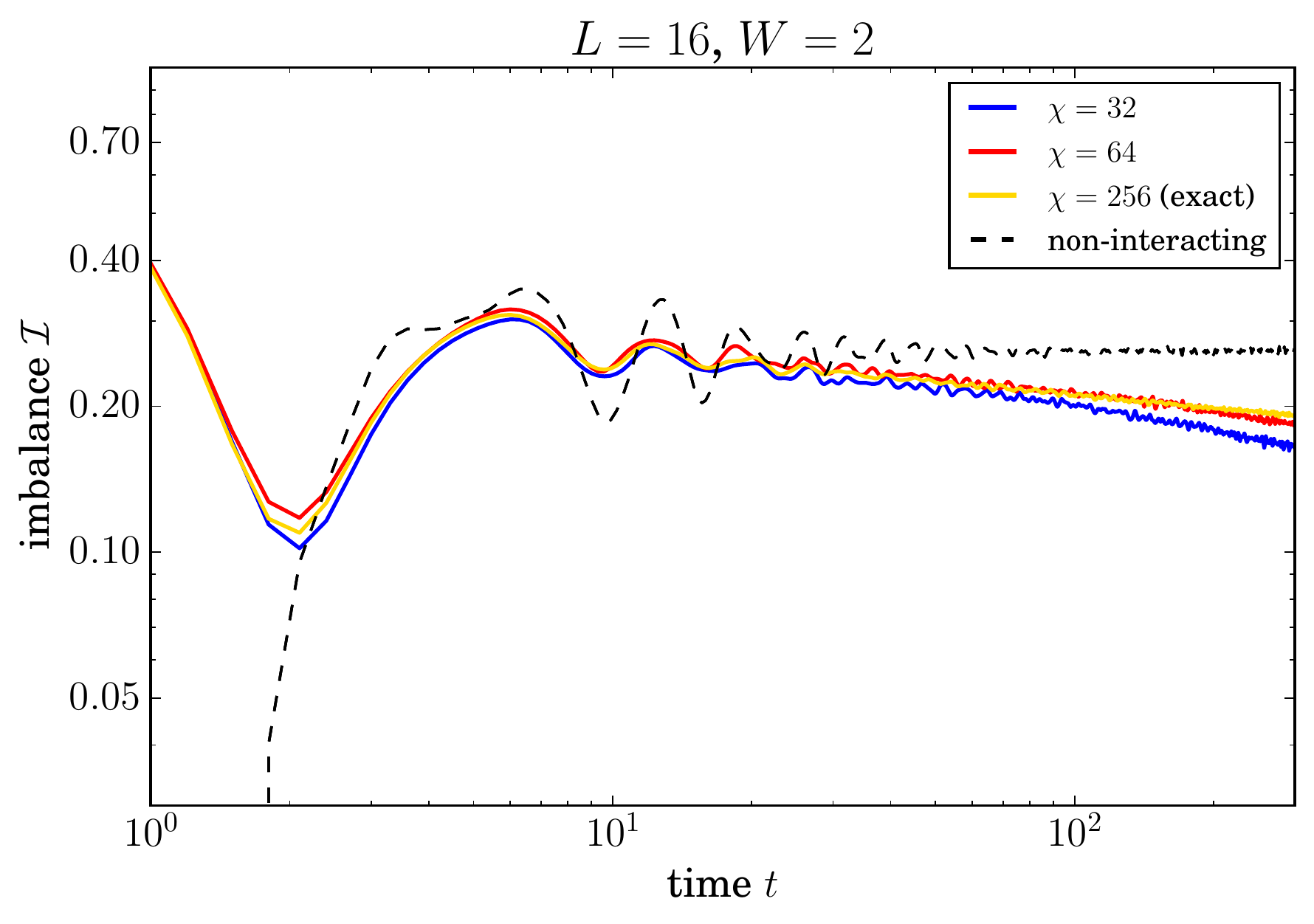}
\caption{
Time evolution of the disorder-averaged imbalance \eqref{eq:imba} as computed by the TDVP with a truncated bond dimension $\chi = 32, 64$ and by exact numerics $\chi = 256$ for a small system $L=16$, up to $t=300$. We consider several hundred independent realizations of disorder. For the sake of comparison, also the non-interacting case $\Delta = 0$ is shown.
}
\label{fig:BenchL16}
\end{figure}



\begin{figure*}[!thb]
\includegraphics[width=0.32\textwidth]{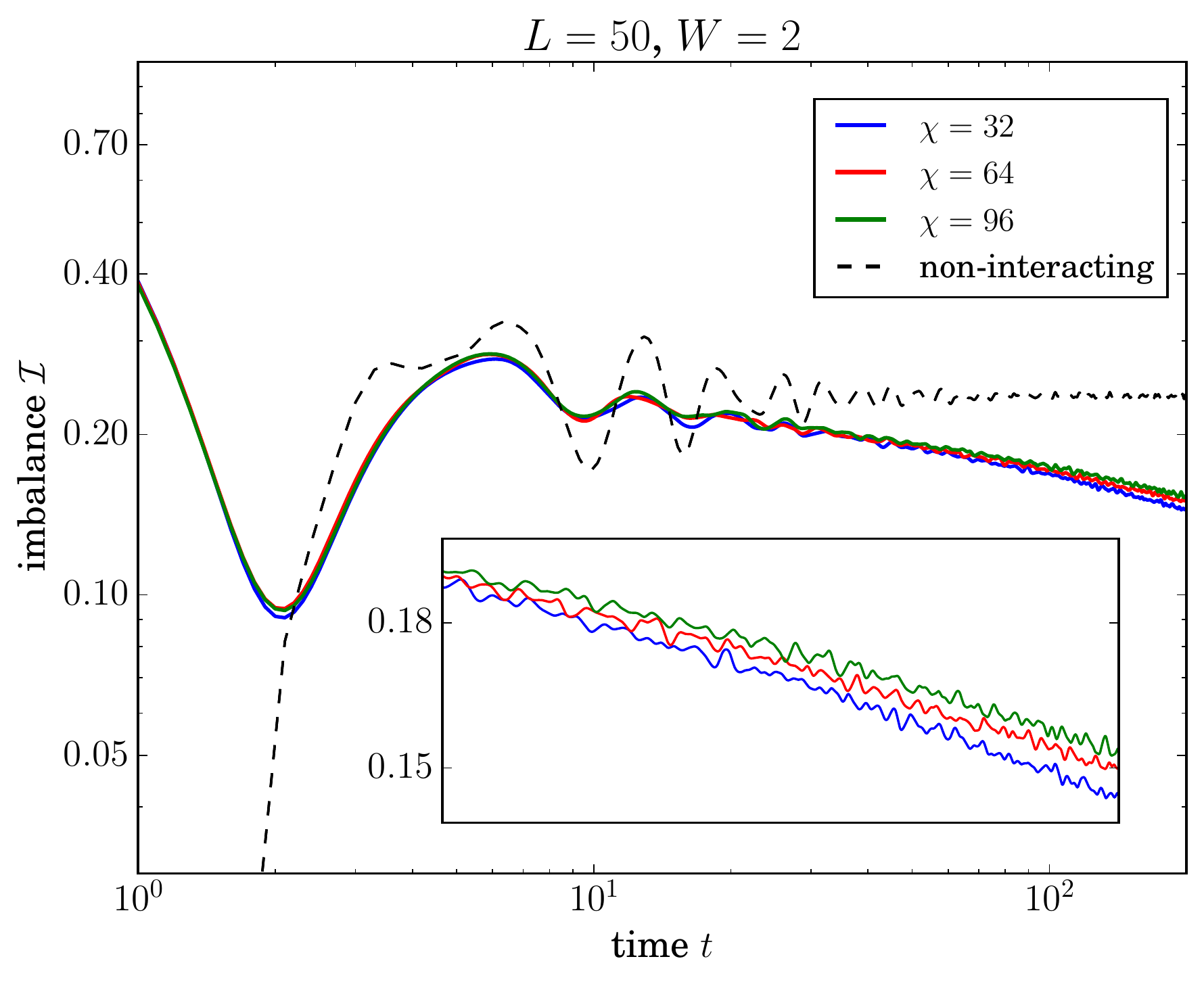}
\includegraphics[width=0.32\textwidth]{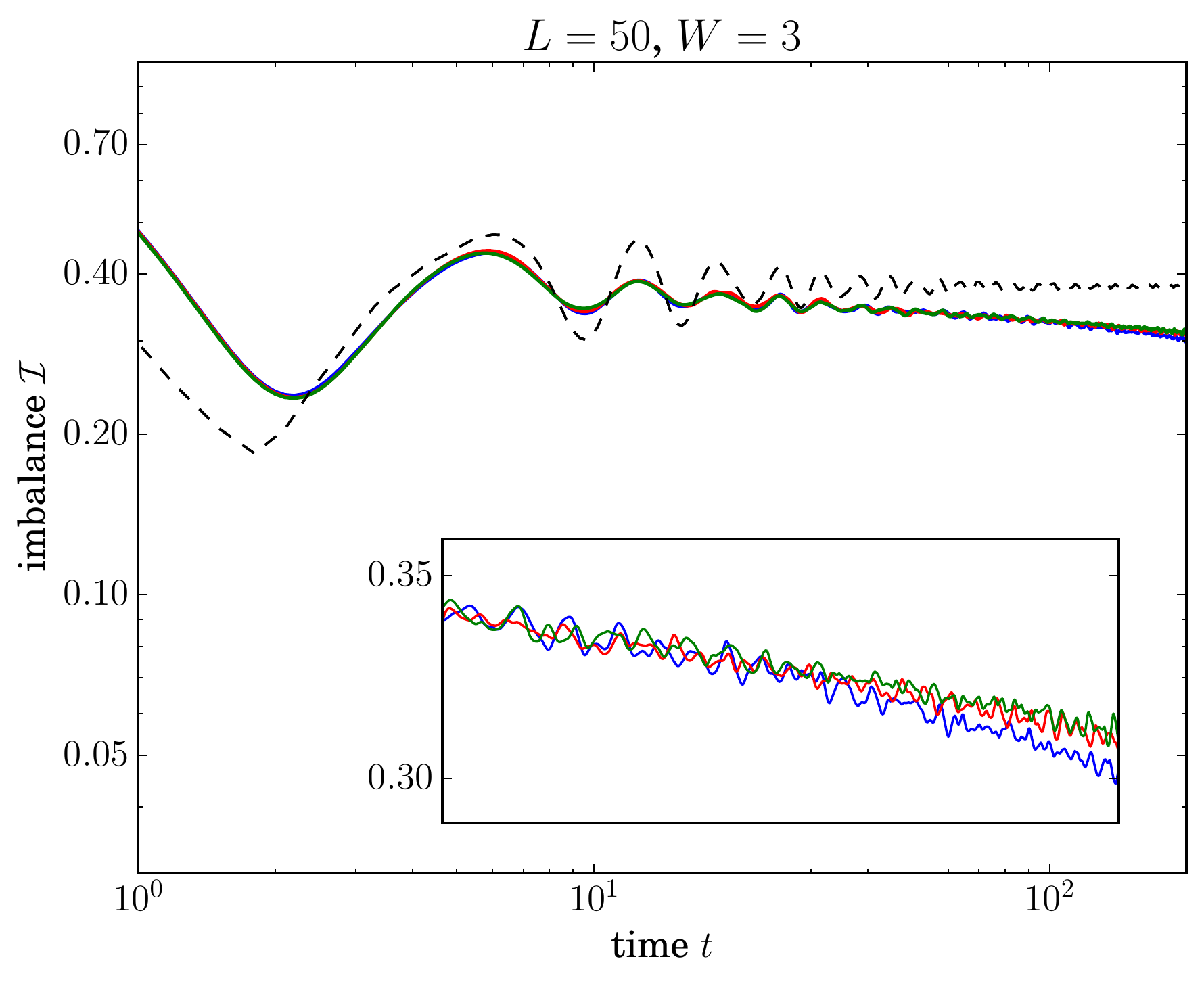}
\includegraphics[width=0.32\textwidth]{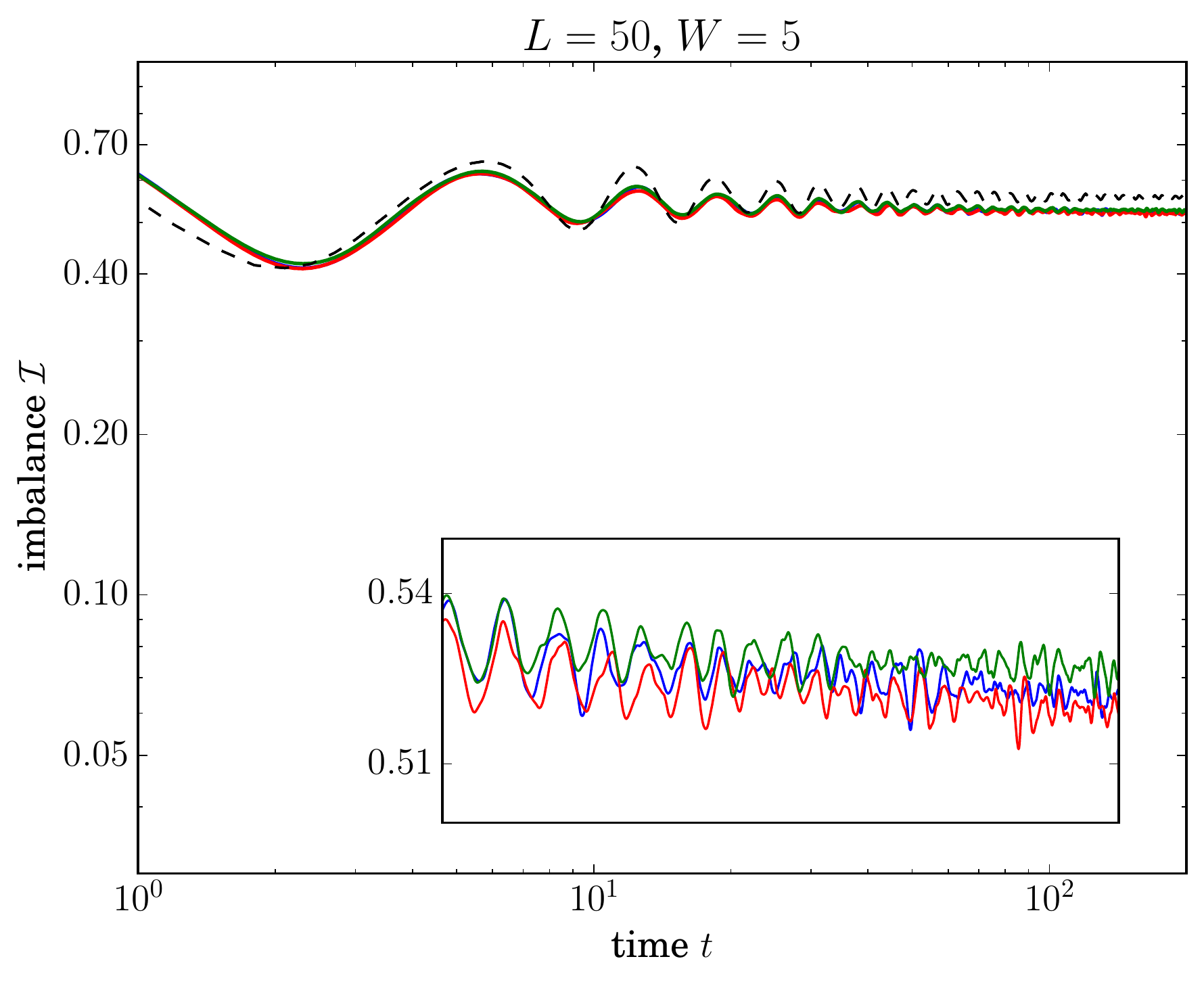}
\caption{
Time evolution of the disorder-averaged imbalance \eqref{eq:imba} as computed by the TDVP with bond dimension $\chi = 32, 64, 96$ for a moderately large system $L=50$ up to $t=200$ and choices of disorder strength $W = 2,3,5$. For comparison, also the non-interacting case $\Delta = 0$ is shown. The fitted coefficients for the power law decay of the imbalance in the window $t \in [100, 200]$ for $W = 5$ are $\beta(\chi = 32) = 0.025 \pm 0.005$, $\beta(\chi = 64) = 0.020 \pm 0.005$, $\beta(\chi = 96) = 0.020 \pm 0.004$, which is in good agreement (within error bars) with the value found for $t \in [50, 100]$ as presented in the main text. Insets show a zoomed region in the time window $t \in [50, 200]$ on a log-log scale.
}
\label{fig:BenchL50}
\end{figure*}


\begin{acknowledgments}

{We thank F.\ Alet, Y.\ Bar Lev, F. \ Evers, M.\ H.\ Fischer, S.\ Goto, C.\ Karrasch, N.\ Laflorencie, D.\ Luitz, S.\ R.\ Manmana, 
M.\ M\"uller, R.\ M.\ Nandkishore, A.\ Scardicchio, B.\ I.\ Shklovskii, and M. \v{Z}nidari\v{c} 
for useful discussions and W.\ Buijsman for pointing out a typographical error in an earlier version of the manuscript.}
TDVP simulations were performed using a Python script based on the open-source {\small evoMPS} library \cite{Milsted2013a}, implementing the single-site algorithm of Ref.\ \cite{Haegeman2016a}.
We are grateful to A.\ Milsted for help with implementing {\small evoMPS}.
TEBD simulations were performed using Open Source Matrix Product States \cite{Wall2012a, Jaschke2018a}.
Simulations for the non-interacting model were performed using the NumPy implementation of {\small LAPACK} \cite{Lapack1999}.
Machine learning algorithms were implemented using {\small TensorFlow} \cite{tensorflow}.
We used Matplotlib \cite{Matplotlib} to generate figures and {\small GNU} Parallel \cite{gnuparallel} for running parallel simulations.
The authors acknowledge support by the state of Baden-W\"urttemberg through bwHPC.
The work of KST, IVG, and ADM was supported by the Russian Science Foundation (Grant No. 14-42-00044).
FS and TN acknowledge support from the Swiss National Science Foundation (grant number: 200021\_169061) and from the European Union's Horizon 2020 research and innovation program (ERC-StG-Neupert-757867-PARATOP).
KST acknowledges support by the Alexander von Humboldt Foundation.

\end{acknowledgments}

\appendix

\section{{Numerical details}} \label{sec:appendix}

Since the TDVP truncates the entanglement spectrum, it is of interest to investigate the convergence with respect to bond dimension \cite{Kloss2017a}. We perform {four distinct} benchmarks. 


\begin{figure*}[!htb]
\includegraphics[width=0.47\textwidth]{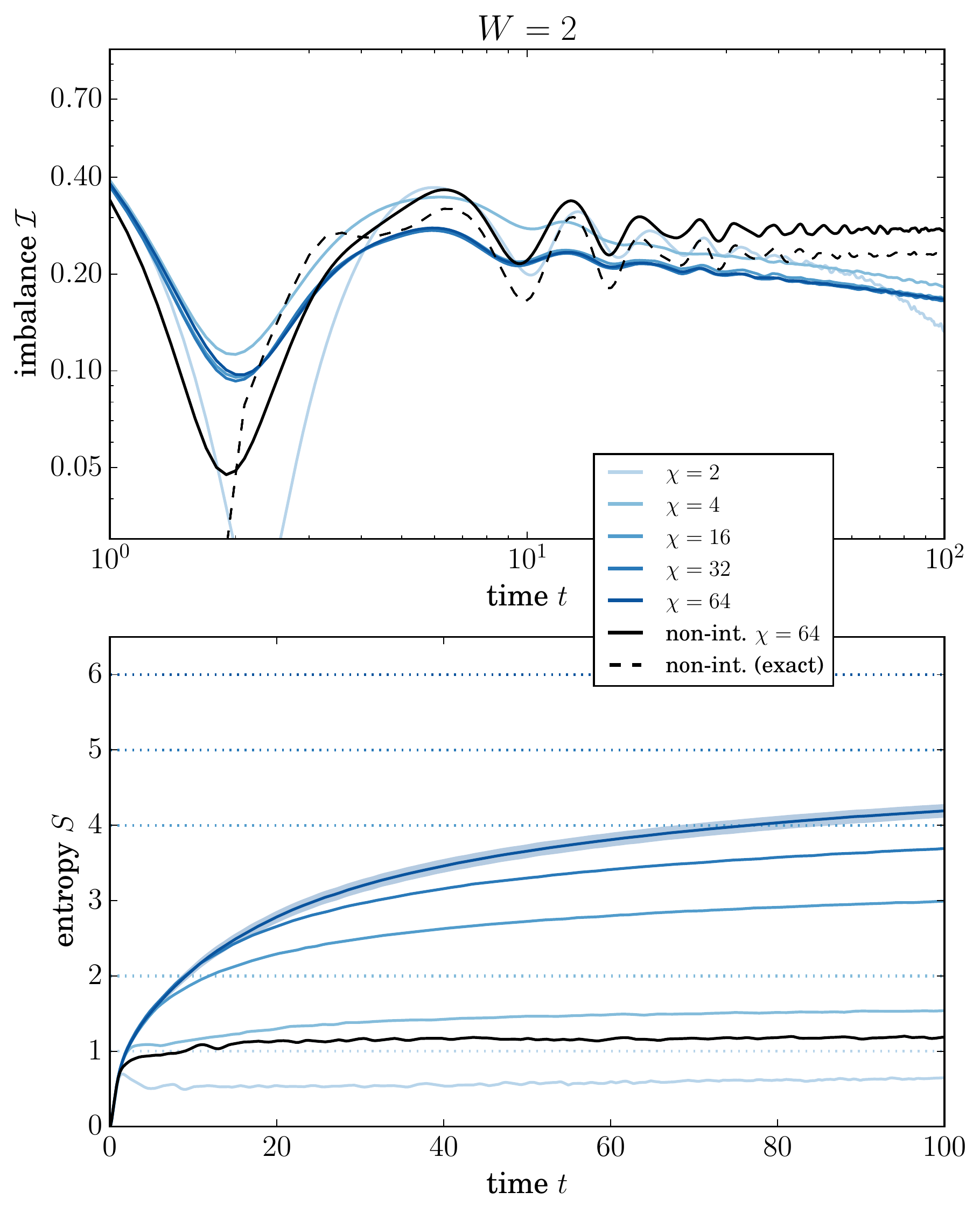}
\includegraphics[width=0.463\textwidth]{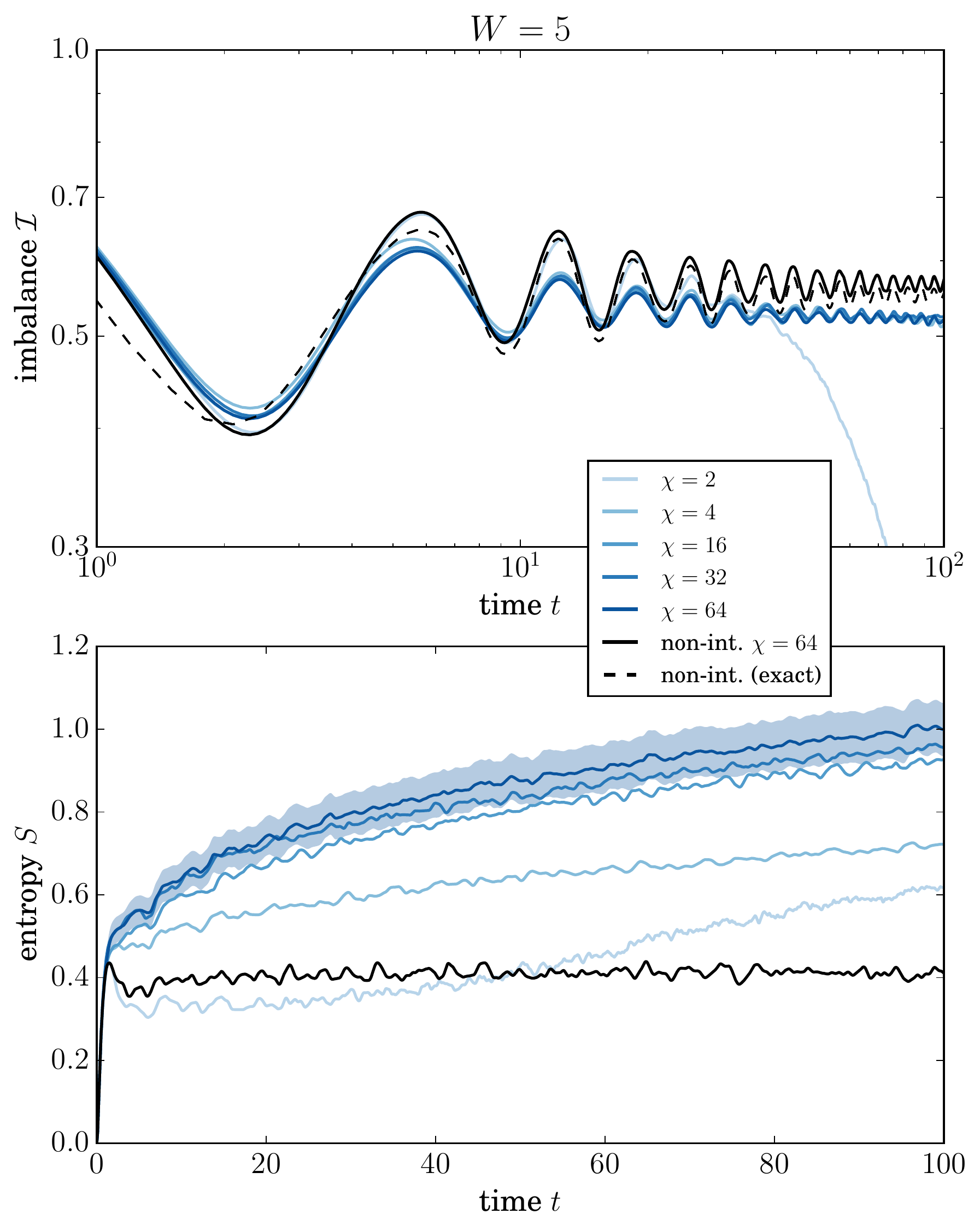}
\caption{ {
\textbf{Top}: time evolution of the disorder-averaged imbalance \eqref{eq:imba} in the case of $L=100$, $W = 2$ (left) and $W = 5$ (right), for various choices of the bond dimension $\chi$. We consider several hundred independent realizations of disorder. For the sake of comparison, also the non-interacting case $\Delta = 0$ is shown, computed both using the TDVP as well as using exact diagonalization.
\textbf{Bottom}: time evolution of the disorder-averaged entropy for the same parameter choices. The horizontal dashed lines indicate the cutoff of the entropy $S_\mathrm{cutoff} = \log_2 \chi$. The shaded region indicates the error ($2\sigma$ intervals) in the average entropy computed from the standard deviation of the data for the case $\chi = 64$.
}
}
\label{fig:BenchW5}
\end{figure*}

\begin{figure}[!ht]
\includegraphics[width=\columnwidth]{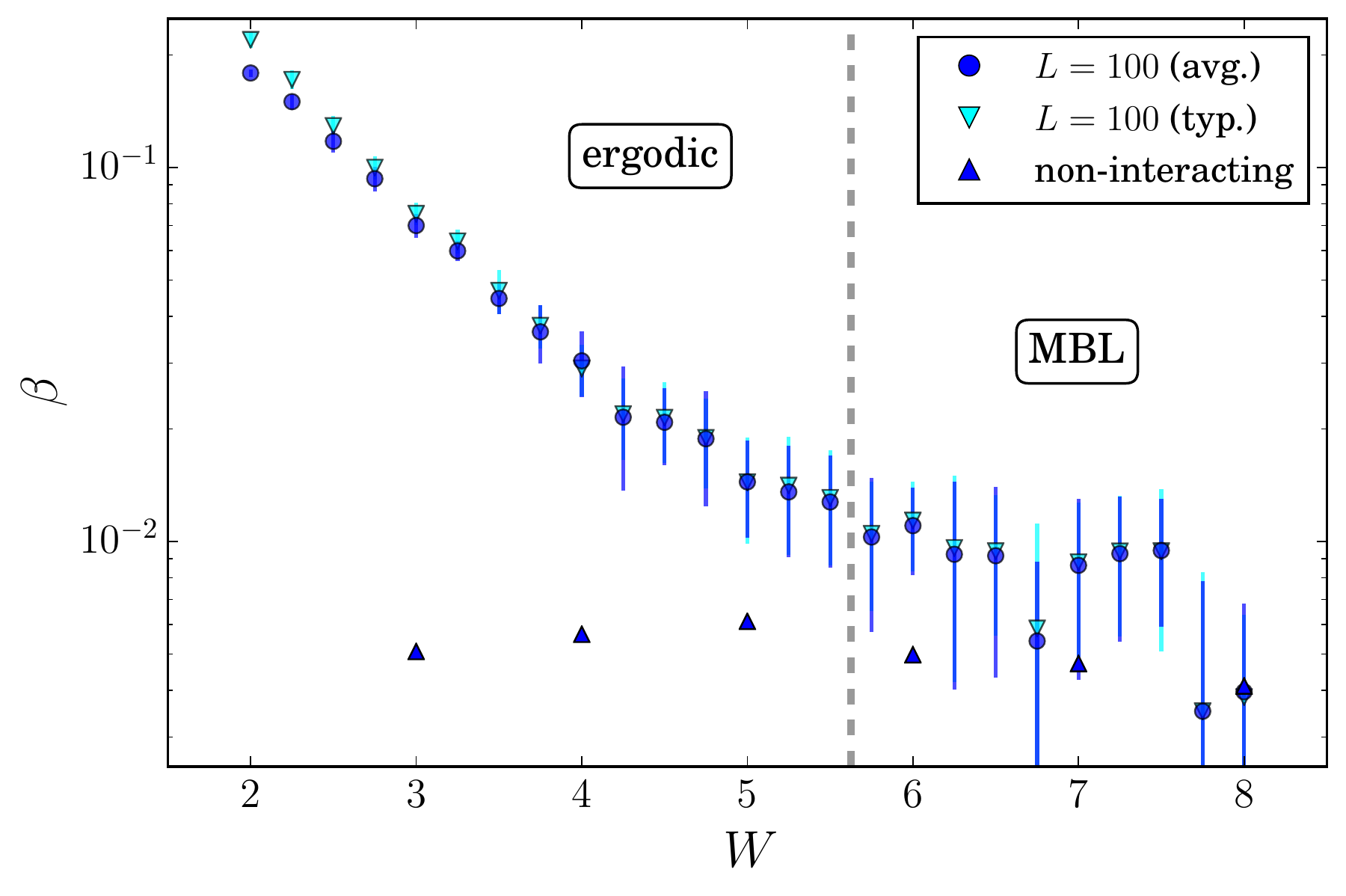}
\caption{ {
Power-law exponent $\beta$ corresponding to the decay of the imbalance 
 $\mathcal{I} \propto t^{-\beta}$ over the window $t \in [50, 100]$, for $L = 100$. Both the average and typical values are considered. Triangles indicate a weak finite-time decay of $\mathcal{I}_\mathrm{A}$ for the non-interacting case $\Delta = 0$, $L=100$. Error bars are $2\sigma$-intervals based on a bootstrapping procedure. The vertical gray line indicates the putative transition from the ergodic to many-body localized regime based on the vanishing of $\beta$ within $2\sigma$, subtracting the systematic error.
}
}
\label{fig:Benchtyp}
\end{figure}

{First, we compare the time evolution of the imbalance obtained within the TDVP approach to numerically exact simulations for a small system $L=16$. Here, we consider the disorder average of independent disorder samples, since we are interested in convergence at the level of this average, rather than at the level of individual realizations. The exact numerics are obtained simply by performing TDVP time evolution with an unrestricted bond dimension. For a system of size $L=16$ an MPS with maximum bond dimension $\chi = 256$ captures the time evolution exactly (for a system of size $L$, non-truncated MPS have a maximum bond dimension of $2^{L/2}$ in the center of the chain). The result is shown in Fig.~\ref{fig:BenchL16}, which shows the time evolution of the initial state up to times $t=300$. Recall that in the main body of the paper, we considered times only up to $t=100$. We observe that even for the weakest disorder considered in the paper ($W=2$), the evolution of imbalance at $\chi=64$ is essentially indistinguishable from the exact evolution in the time window of interest.}

{Second, we consider a larger system $L=50$ and analyze whether the result converges with bond dimension, where we consider times up to $t=200$ and bond dimensions up to $\chi = 96$, higher than in the main body of the paper. The result is shown in Fig.~\ref{fig:BenchL50}. The difference between the values of $\beta$ extracted from the curves for $\chi=64$ and $\chi=96$ is within the statistical error bars of Fig.~\ref{fig:Benchtyp} for all three values of $W$ (recall that different disorder realization were used for different bond dimensions). }

{Third, we consider the effect of \emph{reducing} the bond dimension in long chains ($L=100$) for the ``worst case'' of the weakest disorder used in the paper ($W=2$) and for relatively strong disorder ($W = 5$) that corresponds to the vicinity of the MBL transition. We also present the TDVP results for the non-interacting case $\Delta = 0$ (in the main body of the paper, the results presented for $\Delta = 0$ are obtained by exact diagonalization). In addition, we show the entanglement entropy as a function of time, to illustrate that the truncation of the entanglement spectrum for our choice of the bond dimension is only relevant deeper into the ergodic regime. These results are shown in Fig.~\ref{fig:BenchW5}. We observe that reducing the bond dimension both at $W = 2$ and at $W=5$ from $\chi=64$ even to $\chi=16$ does not significantly affect the evolution of the imbalance, with the corresponding values of $\beta$ again falling within the statistical error bars. The entropy itself is more sensitive to the bond dimension, which is related to the broad distribution of this quantity, see Fig.~\ref{fig:enttime}b in the main text. Nevertheless, we find agreement within error bars (see Fig.~\ref{fig:BenchW5}, lower panels) in the case $W = 5$ between $\chi = 32$ and $\chi = 64$, so that in this case even the entanglement entropy has converged with bond dimension, and the entropy never gets close to the cutoff value $\log_2 \chi$. This is in contrast to the case $W = 2$, where convergence is observed in the imbalance, but not the entropy, where clear cutoff effects occur (dashed lines in the lower left panel).} 

{As a final benchmark, the TDVP simulation of the non-interacting chain ($\Delta=0$) at $W=2$ and $W=5$ does not show any delocalization trend within our time interval (see Fig.\ \ref{fig:BenchW5}, black lines). This should be contrasted with the TDVP result of Ref.~\cite{Kloss2017a} obtained for the XX model at weaker disorder ($W=1$), where the truncation at $\chi=64$ introduced diffusion-like departure from Anderson localization already for $t<100$. Hence, in our model, for $W \geq 2$ (and possibly smaller $W$), there is no spurious dephasing induced by the entanglement truncation procedure (at least, within our time window).}

{From these approaches, we conclude that for the times we consider (up to $t=100$), a bond dimension of $\chi = 64$ provides an excellent approximation for not too weak disorder $W \geq 2$; an approximation that only improves with increasing disorder. This conclusion is in agreement with the findings of Ref.~\cite{Kloss2017a}
concerning the reliability of the TDVP approach for studying transport properties of a disordered XXZ model. In that paper, the short- and long-time behavior of the subdiffusive spreading of spin configurations 
agreed very well for the range of bond dimensions $32 \leq \chi \leq 128$ for even weaker disorder ($W=1.5$) than used in our work ($W\geq 2$).}

Our conclusion is further corroborated by comparison with a completely independent implementation of the time-evolving block decimation method, for which we compute the power law decay coefficient $\beta$ for $W = 5$, finding excellent agreement with the TDVP (see Fig.\ \ref{fig:beta}). Therefore, our numerical results are certainly accurate in the disorder and time window we have considered in this work. Moreover, in the moderately strongly disordered case $W = 5$ we find good agreement between power laws obtained in the window $t \in [50,100]$ and those in the window $t \in [100,200]$, providing evidence for the survival of these power laws in the long-time limit. {In addition, at this value of $W$ the bond dimension can be substantially reduced before noticeable deviations occur.}

{Finally, in addition to benchmarking TDVP results, we present a more detailed look at the power-law coefficient $\beta$ as determined from the decay of the imbalance. In Fig.~\ref{fig:Benchtyp} we show $\beta$ as determined from the average as well as the typical (the exponential of the averaged logarithm) imbalance. If these two quantities are equal, then both the average and typical values of $\beta$ are representative.  Deeper in the ergodic regime, these results differ slightly, indicating that the distributions in this regime deviate somewhat from a Gaussian. For $W \gtrsim 3$ the results overlap, showing that the average and typical values are well-behaved closer to the transition. In Fig.~\ref{fig:Benchtyp} we also mark the bound for the ergodic phase based on the prescription described in the main text.}

{Summarizing, the above benchmarks demonstrate that our numerical results are reliable in the disorder and time windows we have considered in the main text, and the approximate nature of the TDVP deeper in the subdiffusive regime certainly does not affect the reliability of our estimate of the extent of the ergodic region as determined by a vanishing $\beta$.}

{\emph{Note --- } while finalizing the manuscript, a study of the applicability and reliability of the one-site and two-site TDVP algorithms in various other systems appeared \cite{Goto2018a}. In that work, it is shown that the one-site algorithm is typically better suited for the time evolution after a quench in non-integrable models, even in the case of relatively strong entanglement growth, as in the models studied therein (where the entanglement growth is faster than in the case of our moderately to strongly disordered system). In the language of Ref.\ \cite{Goto2018a}, our implementation consists of a hybrid approach, where the initial product state's MPS manifold, with $\chi = 1$, is quickly expanded at each time step, after which we use the one-site algorithm with a fixed bond dimension.}

\bibliography{ref}

\end{document}